\documentclass[twocolumn,english,prl,superscriptaddress]{revtex4}
\usepackage{amsmath,amssymb,amsfonts}
\usepackage{CJK}
\usepackage{bm}
\usepackage{tikz}
\usepackage{pgfplots}
\usetikzlibrary{matrix,fit}
\usepackage{graphicx} 
\usepackage{braket}
\usepackage{subfig}
\usepackage{caption}
\usepackage{mathrsfs}
\usepackage{float}
\usepackage[utf8]{inputenc}
\usepackage{pgflibraryarrows}
\usepackage{lipsum} 
\usepackage{pgflibrarysnakes}

\begin{document}
\renewcommand{\raggedright}{\leftskip=0pt \rightskip=0pt plus 0cm}
\captionsetup[figure]{name={FIG.},labelsep=period}
\title{Visualization of higher-order topological insulating phases in two-dimensional dielectric photonic crystals}

\author{Bi-Ye Xie}\thanks{These authors contributed equally to this work.}
\affiliation{National Laboratory of Solid State Microstructures, Nanjing University, Nanjing 210093, China}
\affiliation{Department of Materials Science and Engineering, Nanjing University, Nanjing 210093, China}

\author{Guang-Xu Su}\thanks{These authors contributed equally to this work.}
\affiliation{National Laboratory of Solid State Microstructures, Nanjing University, Nanjing 210093, China}
\affiliation{School of Physics, Nanjing University, Nanjing 210093, China}

\author{Hong-Fei Wang}\thanks{These authors contributed equally to this work.}
\affiliation{National Laboratory of Solid State Microstructures, Nanjing University, Nanjing 210093, China}
\affiliation{Department of Materials Science and Engineering, Nanjing University, Nanjing 210093, China}
\author{Hai Su}

\affiliation{National Laboratory of Solid State Microstructures, Nanjing University, Nanjing 210093, China}
\affiliation{School of Physics, Nanjing University, Nanjing 210093, China}
\author{Xiao-Peng Shen}
\affiliation{School of Physical Science and Technology, China University of Mining and Technology, Xuzhou, 221116, China}

\author{Peng Zhan}
\email[]{zhanpeng@nju.edu.cn}
\affiliation{National Laboratory of Solid State Microstructures, Nanjing University, Nanjing 210093, China}
\affiliation{School of Physics, Nanjing University, Nanjing 210093, China}
\affiliation{Collaborative Innovation Center of Advanced Microstructures, Nanjing University, Nanjing 210093, China}
  
\author{Ming-Hui Lu}
\email[]{luminghui@nju.edu.cn}

\affiliation{National Laboratory of Solid State Microstructures, Nanjing University, Nanjing 210093, China}
\affiliation{Department of Materials Science and Engineering, Nanjing University, Nanjing 210093, China}
\affiliation{Jiangsu Key Laboratory of Artificial Functional Materials, Nanjing 210093, China}
\affiliation{Collaborative Innovation Center of Advanced Microstructures, Nanjing University, Nanjing 210093, China}
\author{Zhen-Lin Wang}

\affiliation{National Laboratory of Solid State Microstructures, Nanjing University, Nanjing 210093, China}
\affiliation{School of Physics, Nanjing University, Nanjing 210093, China}
\affiliation{Collaborative Innovation Center of Advanced Microstructures, Nanjing University, Nanjing 210093, China}
\author{Yan-Feng Chen}
\affiliation{National Laboratory of Solid State Microstructures, Nanjing University, Nanjing 210093, China}
\affiliation{Department of Materials Science and Engineering, Nanjing University, Nanjing 210093, China}
\affiliation{Collaborative Innovation Center of Advanced Microstructures, Nanjing University, Nanjing 210093, China}

%\pacs{42.70.Qs, 42.25.Bs, 78.20.Bh}

\begin{abstract}
The studies of topological phases of matter have been extended from condensed matter physics to photonic systems, resulting in fascinating designs of robust photonic devices. Recently, higher-order topological insulators (HOTIs) have been investigated as a novel topological phase of matter beyond the conventional bulk-boundary correspondence. Previous studies of HOTIs have been mainly focused on the topological multipole systems with negative coupling between lattice sites. Here we experimentally demonstrate that second-order topological insulating phases without negative coupling can be realized in two-dimensional dielectric photonic crystals (PCs). We visualize both one-dimensional topological edge states and zero-dimensional topological corner states by using near-field scanning technique. To characterize the topological properties of PCs, we define a topological invariant based on the bulk polarizations. Our findings open new research frontiers for searching HOTIs in dielectric PCs and provide a new mechanism for light-manipulating in a hierarchical way.
\end{abstract}
\maketitle

{\it{Introduction}}.--- One of the most enchanting developments of condensed matter physics over the past few decades has been the discovery of topological phases of matter primarily found in electronic systems~\cite{TI1,TI2} and recently extended to bosonic systems such as photonics~\cite{PTI1,PTI2,PTI3,PTI4,PTI5,PTI6,PTI7,PTI8,PTI9,PTI10, d1,d2} and phononics~\cite{PNI1,PNI2,PNI3,PNI4,PNI5,PNI6}. A key feature of the topological insulators is the backscattering-immune edge states which are robust against perturbations and provide potential designs of various topological devices~\cite{PTI1,PTI2,PTI4,PNI3,PNI5,PTI3}. Typically, $n$-dimensional ($n$D) topological insulators (TIs) have $(n-1)$D edge states which is defined as the bulk-boundary correspondence (BBC)~\cite{BBC}. However, a new kind of TIs defined as the higher-order topological insulators (HOTIs), have been recently proposed in tight-binding models in electronic systems which go beyond the traditional BBC description~\cite{HOTI1,HOTI2,HOTI3,HOTI4,HOTI5,HOTI6,HOTI7,HOTI8,HOTI9,HOTI10,HOTI11,HOTI12,HOTI13,HOTI14,HOTI15}. Concretely, the $m$th-order TIs have $n$D gapped bulk states and $(n-1)$D, $(n-2)$D, ..., $(n-m-1)$D gapped edge states while having $(n-m)$D gapless edge states. The arising of these lower-dimensional topological edge states can either stem from the quantization of quadrupole moments such as the topological quadrupole insulators~\cite{HOTI1} which have been realized in mechanics~\cite{HOTI2}, microwave systems~\cite{HOTI3} and topolectrical circuits~\cite{HOTI4}, or stem from the quantization of the dipole moments ~\cite{HOTI1} such as the HOTIs in 2D breathing kagome lattice~\cite{HOTI8} which have been realized in sonic crystals~\cite{HOTI12,HOTI13,HOTI14} and a waveguide array~\cite{HOTI11}. 

Photonic crystals (PCs) offer irreplaceable opportunities to investigate theoretical physics in a highly controllable way~\cite{PC1,PC2}. Moreover, they have a wide range of applications in manipulating the propagation of electro-magnetic (EM) waves and designing novel photonic devices from microwaves to optical waves. In particular, dielectric PCs which have no undesirable Ohmic-loss effect compared to those with metallic and plasmonic materials play an important role in realizing photonic integrated circuits and resonator antennas with high radiation efficiency~\cite{PC3,PC4}. Previously, topological insulating phases in dielectric PCs have been experimentally demonstrated with traditional BBC (the first-order TIs)~\cite{PC4}.
A recent theoretical study shows that it is possible to realize HOTIs in 2D pure-dielectric PCs~\cite{HOTI9}. However, the experimental realizations of HOTIs in pure-dielectric PCs without negative coupling are still challenging and waiting to be achieved.

In this Letter, we report on the experimental realization of a second-order topological insulator (SOTI) in a 2D dielectric PC which is 2D photonic extension of Su-Schrieffer-Heeger (SSH) lattice~\cite{SSH1, SSH2,SSH3, SSH4}. Instead of introducing a negative coupling which is essential to the formation of topological quadrupole insulators, we demonstrate that the dipole moments quantized by mirror symmetries will lead to second-order topological insulating phases. We engineer a square meta-structure formed by two pieces of PCs where the inner PC is in a topological non-trivial phase while the outer PC is in a topological trivial phase. By applying the near-field scanning technique, we visualize both the first-order and the second-order topological insulating phases in the same structure which can be simply controlled by the geometric parameters, implying a hierarchical structure in topological insulating phases~\cite{HOTI15}.

{\it{Crystal structure and bulk band structure}}.---Our 2D photonic crystals (PCs) possess a square lattice geometry with four artificial atoms formed in a unit cell as depicted in Fig. 1(a). There are two competing parameters in a PC: the intra-cell distance of nearest atoms $d_1$ and the inter-cell distance of nearest atoms $d_2$. $d_1$ and $d_2$ modulate the coupling strengths between nearest lattice sites and we notice that $d_1+d_2=a$ where $a$ is the lattice constant. For simplicity, we define $\Delta=d_1-d_2$ which determines the lattice structure of a PC if other parameters are fixed. In this Letter, we consider PCs with a finite height in $z$-direction and set the height of cylinders $h=13.5mm$, lattice constant $a=20mm$ and radius of the cylinders $r=2.4mm$ as shown in Fig. 1(c).

Next, we show the evolution of band structures for transverse magnetic (TM) modes by numerically simulating our PCs. In all simulations, we consider perfect electric conductor (PEC) boundaries in the upper and lower plates perpendicular to $z$ direction as shown in Fig. 1(c) and set the relative dielectric permittivity $\epsilon=6.1$. The above implementations ensure that the 2D approximation is valid. We start from a simple case where $\Delta=0$ (indicated by grey circles in Fig. 1(b)) and the PC is a conventional 2D square lattice PC (with one atom in unit cell). In this case, the first band and the second band are degenerate at $X$ point as shown in Fig. 1(d) (middle panel). Next, we consider two cases: $\Delta=-0.11a$ which corresponds to the “shrunken" lattice (indicated by dashed blue circles in Fig. 1(b)) and $\Delta=0.11a$ which corresponds to the “expanded" lattice (indicated by dashed red circles in Fig. 1(b)). In both cases, the previous gapless point is opened and a full photonic band gap (PBG) emerges. The band structures for these three cases are presented in Fig. 1(d). We notice that the “shrunken" lattice and “expanded" lattice are in distinct topological phases which are connected by a band-inversion process. To demonstrate it clearly, we calculate the electric field distribution of the unit cell for two cases as shown in Fig. 1(e). Due to the inversion symmetry, the parities of the first band and the second band can be defined as the eigenvalues of inversion operator. As shown in of Fig. 1(d) and Fig. 1(e), the parities for the first and the second band have changed as we change $\Delta=-0.11a$ to $\Delta=0.11a$, indicating the bands have been inverse.

\begin{figure}
\centering
\includegraphics[scale=0.19]{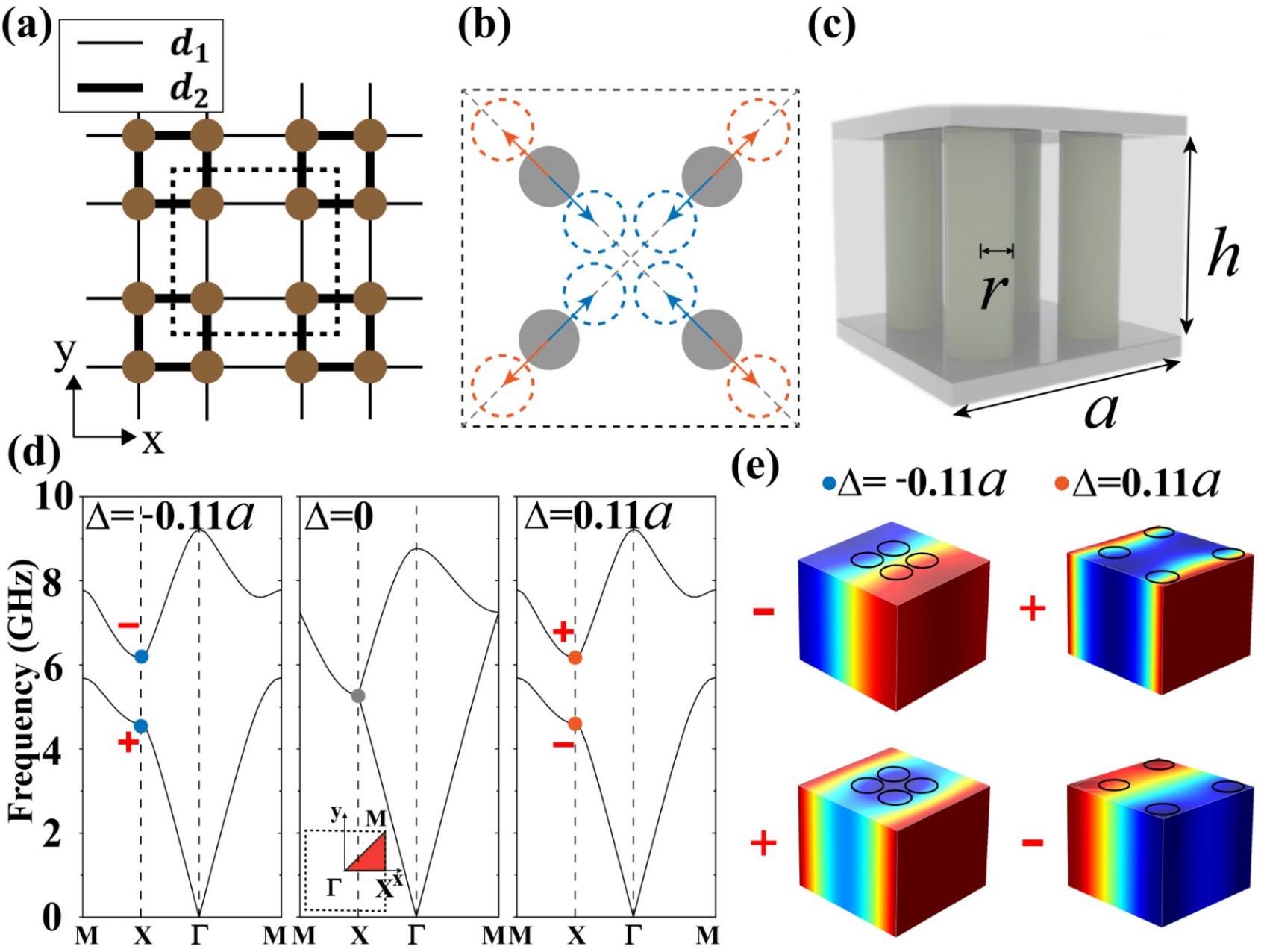}
\captionsetup{format=plain, justification=raggedright}
\caption{Crystal structures and bulk band structures. (a) 2D lattice structure of a PC with $d_1$ and $d_2$ representing intra-cell distance and inter-cell distance of the nearest atoms. (b) Different configurations of PCs. The dashed blue (red) circles represent “shrunken” (“expanded”) configuration and gray circles represent “normal” square lattice without tetramerization. (c) 3D structure of a unit cell. The upper and lower plates represent metallic surfaces which confine the electro-magnetic fields to form a 2D PC. (d) Band structures for $\Delta=-0.11a$ (left), $\Delta=0$ (middle) and $\Delta=0.11a$ (right) which represent “shrunken”, “normal” and “expanded” lattice respectively. We notice that for “shrunken” and “expanded” lattice, the dispersion of band structures are the same but with an inversed parities of the first and second bands at $X$ point (denoted by ‘+’ and ’-’ symbols). (e) The electric field distributions in a unit cell of the first band (lower panels) and the second band (upper panels) at $X$ points in the first Brillouin zone. The odd and even parities can be directly seen from the field distributions.}
\label{fig:1}
\end{figure} 

{\it{Topological phase transition induced by deformation of lattice structure}}.---Here we investigate the relation between the size of the PBG and $\Delta$. The result (phase diagram) is given in Fig. 2(a). It is intriguing to notice that there is a metallic phase other than two insulating phases for $\Delta\neq 0$. Although the local gapless point at $X$ point is opened, the PBG in the first Brillouin zone does not appear when $|\Delta|$ is relatively small. The phase diagram is similar to the theoretical tight-binding (TB) model (2D SSH model)~\cite{HOTI9, SSH3}. From the perspective of TB approximation , when $\Delta$ is non-zero, a tetramerized configuration of the nearest-neighbor (NN) coupling is formed and the mass term in the effective Hamiltonian discribing the low-energy physics around the gapless point is non-zero. We found two areas of fully gapped phases when $\Delta<-0.087a$ and $\Delta>0.087a$ which are represented by blue and red colors in Fig. 2(a) respectively. These two areas have two mass terms with different signs respectively and hence are topological distinct to each other. Since there is no gap-closing process among each individual area, the band structures can be connected by adiabatic evolutions and thus belong to the same topological class. Detailed discussions about the topological equivalence between the band structures of our PCs and those of 2D SSH model are provided in Supplementary Materials~\cite{SUP}.

To characterize the topological properties of our PCs, we define a topological invariant based on the 2D polarization. The topological phases can be characterized by 
\begin{equation}
P_i=-\frac{1}{(2\pi)^2}\int_{BZ} d^2\bm{k}\mathrm{Tr} [\hat{{\cal A}}_i] , \quad i=x,y
\end{equation}
where $(\hat{{\cal A}}_i)_{mn}(\textbf{k})=\mathrm{i}\bra{u_m(\textbf{k})}\partial_{k_i}\ket{u_n(\textbf{k})}$, with $m$, $n$ run over all bands below the PBG. $\ket{u_m(\textbf{k})}$ is the periodic part of the electric field for the $m$th band. The 2D polarization is simply related to the 2D Zak phase via $\theta_i=2\pi P_i$ for $i=x,y$. For $\Delta>0.087a$, $\bm{P}=(P_x, P_y)=(\frac{1}{2},\frac{1}{2})$ which implys that the PC is in topological non-trivial insulating phase. Similarly when $\Delta<-0.087a$, $\bm{P}=(P_x, P_y)=(0, 0)$ which implys that the PC is in topological trivial phase.

A remarkable feature of topological insulating phases is that there are often topological edge states appearing at the interface of two distinct topological materials. To demonstrate the existence of edge states, we consider a combined structure of PCs where a topological non-trivial PC (with $\Delta=0.11a$) is jointed by a topological trivial PC (with $\Delta=-0.11a$) as shown in Fig 2(b). The projected band structure is numerically calculated and shown in Fig. 2(b). We use the PEC boundary condition for boundaries parallel to the interface of two PCs while using Floquet periodic boundary condition for boundaries perpendicular to the interface and set other parameters the same as those in Fig. 1. As seen in Fig. 2(b), an interface state (indicated by solid blue line) appears in the middle of PBG. For comparison, we provide the projected band structure of two trivial PCs (with $\Delta=-0.11a$ and $\Delta=-0.12a$ respectively) as shown in Fig. 2(c) which possesses no edge states. 
The simulated field distribution of the combined structure for the non-trivial case is shown in Fig. 2(d). A 1D localized state emerges and decays fast away from the interface (indicated by the white dashed line). 

\begin{figure}
\centering
\includegraphics[scale=0.2]{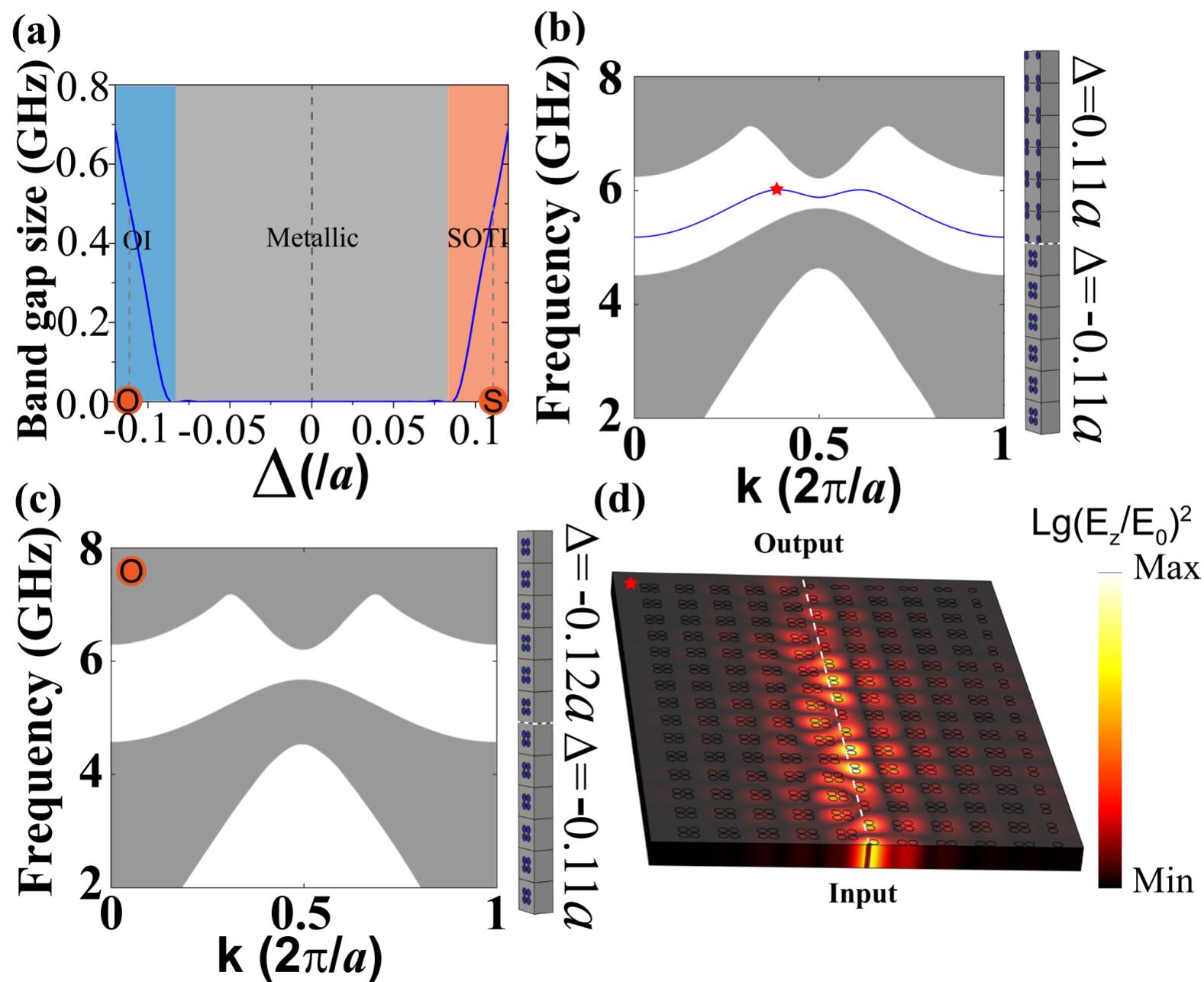}
\captionsetup{format=plain, justification=raggedright }
\caption{Phase diagram and topological edge states. (a) The existing ordinary insulator (OI) phase, metallic phase and second-order topological insulator (SOTI) phase in PCs with respect to different parameters $\Delta$. (b) The projected band structure of two combined PCs within different topological phases. A photonic band gap (PBG) with 1D edge state is presented. (c) The projected band structure of two combined PCs within the same topological phases. A PBG with no edge state is presented. (d) A simulated edge state with a frequency at 6.00 GHz (represented by a red star in (b)).}
\label{fig:4}
\end{figure}

{\it{Experimental observation of corner states in a meta-structure}}.---We now extend the situation from previous combined structure to a square meta-structure as shown in Fig. 3(a). The meta-structure is realized by placing a topological non-trivial PC (denoted as SOTI) with $10\times10$ periods, surrounding by four-layer trivial PC (denoted as OI (ordinary insulator)). $\Delta$ for SOTI and OI are $0.11a$ and $-0.11a$ respectively with other geometric parameters the same as those in Fig. 1. The SOTI and OI are distinct in topology, and their PBGs overlap with each other. Fig. 3(b) shows the eigenmodes of the meta-structure at frequencies between 5.50 GHz to 6.50 GHz. The numerical calculation with Floquet periodic boundary condition for boundaries in $x$ and $y$ directions indicates that four near-degenerate mid-gap states (indicated by blue dots) emerge at the PBG at frequencies around 6.25 GHz between the 1D edge states (indicated by yellow dots) and the bulk bands (indicated by grey dots). The electric field distribution from the calculation is shown in Fig. 3(c) which clearly demonstrates that these four states are strongly localized at four corners of the box-shape boundary (indicated by white dashed line) in the meta-structure.

To experimentally observe corner states, we fabricate a meta-structure of PCs consisting of alumina cylinders with relative dielectric permittivity $\epsilon=10$. To prevent the TM modes from leaking into the air, we use two flat metallic plates to cover the bottom and top of the PCs, mimicking the PEC boundaries. We then place an excitation source at the upper-right corner of the boundary of two PCs as depicted in Fig. 3(d). By using near-field scanning technology, we obtain quantitatively the distribution of the out of plane electric field along the whole interface and visualize the result obtained at 6.26 GHz as depicted in Fig. 3(d). The field distributions are measured by a coaxial prober-fed and the data acquisitions are measured by a vector network analyzer (Agilent E5063A). From Fig. 3(d), we see a  strong concentration of field distribution in four dielectric cylinders at the corners of meta-structure, indicating the sub-wavelength character of corner states. The equal field strengthes of corner states as shown in  Fig. 3(c)-(d) are due to the quantization of $1/4$ total number of photons at the corners originated from a mismatch between the number of photons to strictly satisfy the $C_4$ crystalline symmetry and the conservation of number of photons injected into the PC which is similar to the quantization of fractional charges of electrons in SSH model and recently proposed $C_n$ symmetric HOTIs~\cite{HOTIC}. We also observe a small difference of frequencies of corner states between simulations and experiments which is induced by an air layer above the PCs (see detailed discussion in Supplementary Materials~\cite{SUP}) and manufacturing accuracy errors of the PCs.

\begin{figure}
\centering
\includegraphics[scale=0.19]{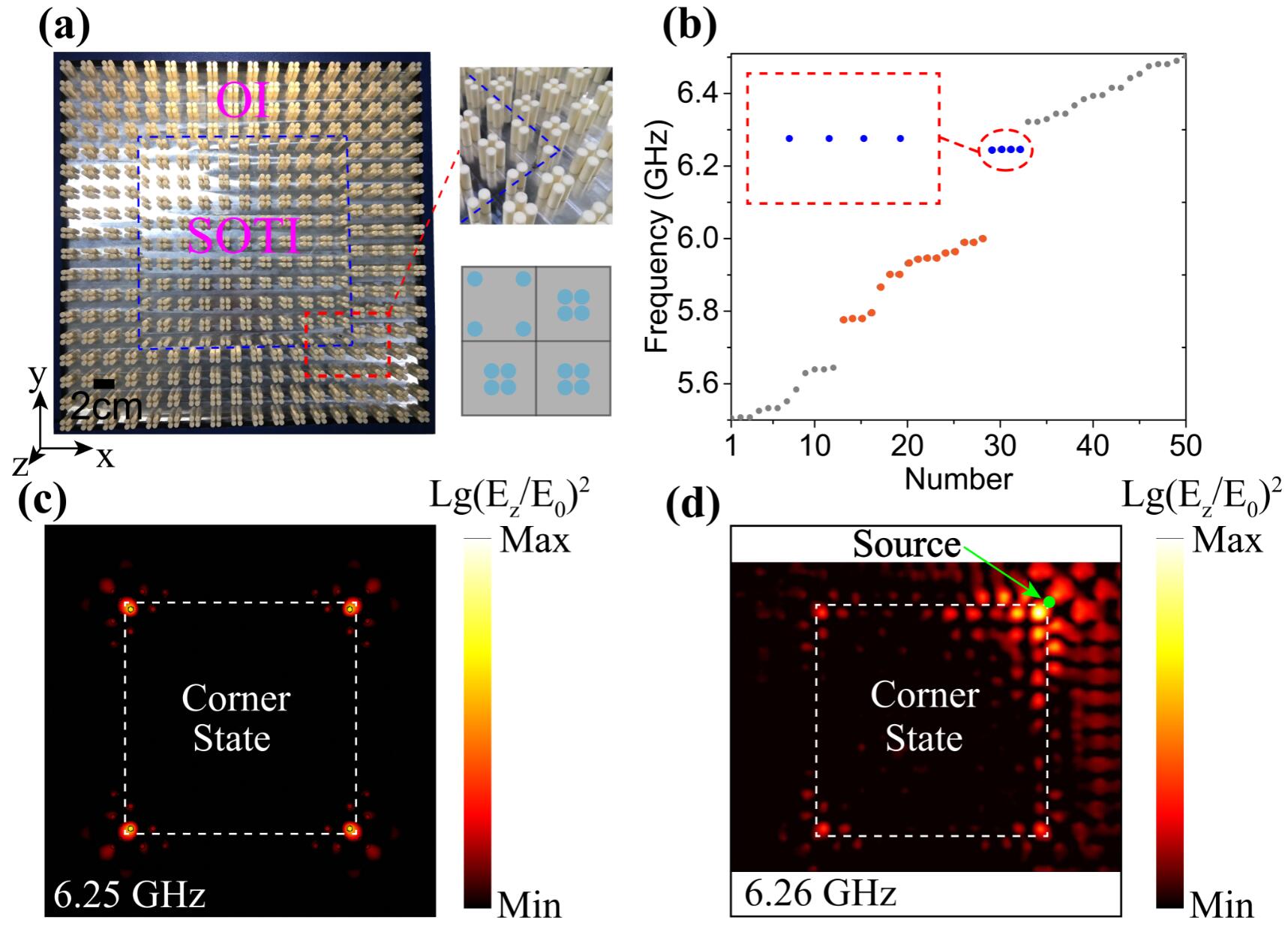}
\captionsetup{format=plain, justification=raggedright }
\caption{Experimental visualization of corner states in meta-structure. (a) Photograph of a meta-structure with upper metallic plate removed, consisting of a SOTI ($\Delta=0.11a$) with $10\times10$ unit cells surrounded by an OI ($\Delta=-0.11a$) with 4 unit cells thickness. The boundary of two PCs is denoted by the blue dashed line. The zoom-in corner structure is presented. (b) Eigenmodes calculation of the meta-structure with the same parameters in Fig. 1(d). Four near-degenerate corner states emerge in the band gap around the frequency of 6.25 GHz as indicated by blue dots. The yellow dots represent the dispersive 1D edge states and grey dots stand for bulk states. (c) A simulated electric field distribution of one of the four corner states. (d) Experimental visualization of corner states at 6.26 GHz. The scanning area is a rectangle instead of a square. The excitation source is placed at the upper-right corner as depicted by the green dot. An enhancement of fields from the source is observed.}
\label{fig:4}
\end{figure}

{\it{Hierarchical topological insulating phases in photonics}}.---To further verify the coexistence of 1D topological edge states and 0D topological corner states, we excite the eigenstates from 5.23 GHz to 6.70 GHz and observe the evolution of the out-of-plane electric field distribution. The same procedure is applied as the previous measurement of corner states. The result is shown in Fig. 4. The normalized density of states (DOS) for bulk, edge and corner points (represented by a grey, yellow and blue dot in the inset respectively) are obtained from the averaged value of field distributions at each point respectively as depicted in Fig. 4 (a). We point out that a non-zero DOS at corner (edge) point can stem form the existence of edge states (bulk states). We find that there is a bulk-edge-corner-bulk evolution as we continuously increase the excitation frequencies. This measurement agrees with the result shown in Fig. 3(b). A small difference in the frequencies between the simulation and experiment is observed due to the air layer above the meta-structure in our experiments. The visualizations of bulk, edge and corner states are presented in Fig. 4(b)-(e) respectively which demonstrate the coexistence of topological insulating phases with the first and second orders, presenting a hierarchical structure of topological insulating phases.

\begin{figure}
\centering
\includegraphics[scale=0.36]{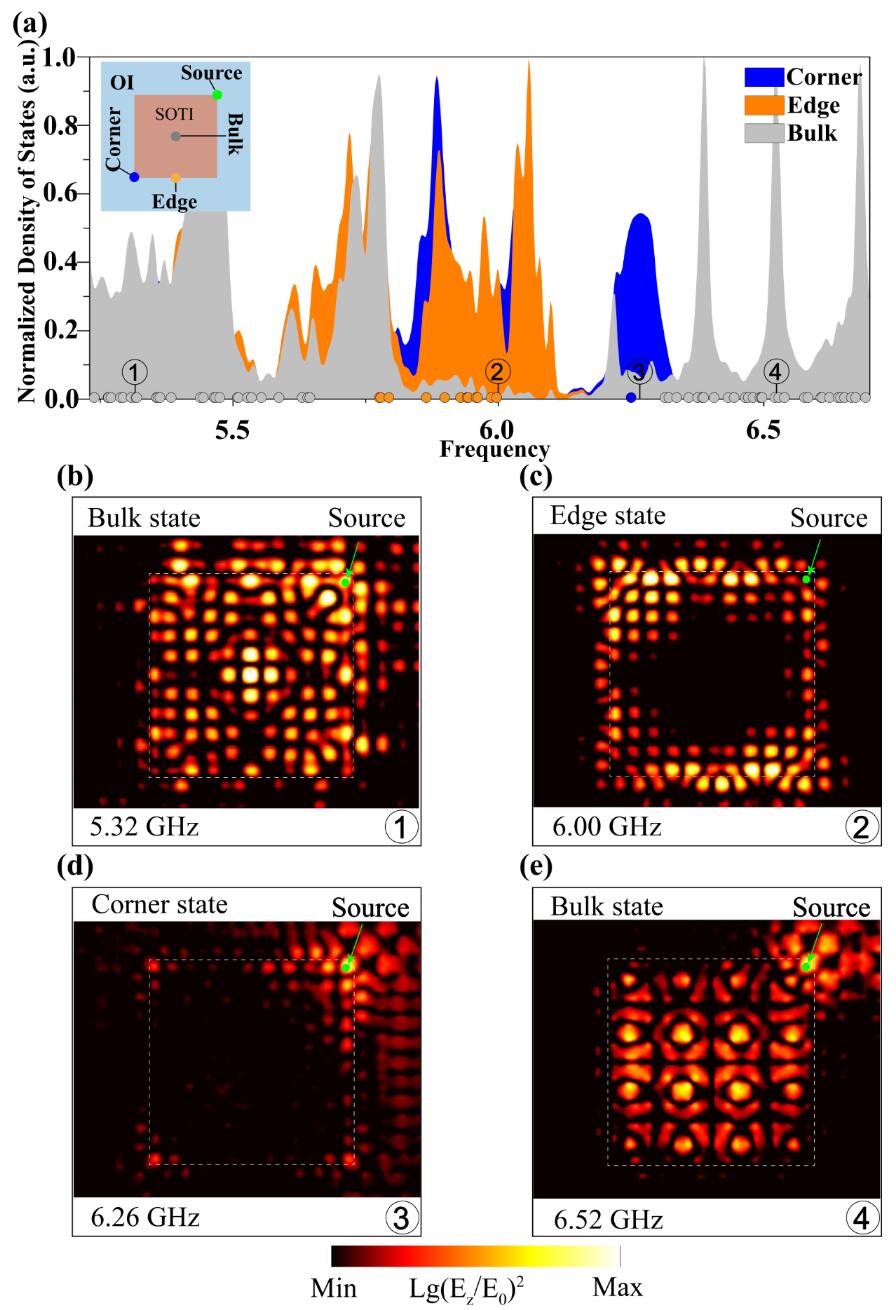}
\captionsetup{format=plain, justification=raggedright }
\caption{Hierarchical structure of topological insulating phases. (a) Normalized density of states for the meta-structure measured at bulk, edge, and corner points as depicted in the inset. The bulk, edge and corner eigenstates from the numerical calculation are represented by dots with grey, yellow and blue colors respectively. Visualization of (b) a bulk state at 5.32 GHz, (c) an edge state at 6.00GHz, (d) a corner state at 6.26 GHz and (e) a bulk state at 6.52 GHz with excitation source placed at the upper-right corner of the boundary (represents by the white dashed line) of two PCs in meta-structure.}
\label{fig:6}
\end{figure}

{\it{Conclusion and disscussion}}.---In summary, we here experimentally demonstrate a 2D SOTI in dielectric PCs and visualize both 1D topological edge states and 0D sub-wavelength corner states. The hierarchical structure of topological insulating phases is observed in a topological non-trivial configuration. These structure-controlled topological phases may pave the way for the realization of 3D photonic HOTIs and higher-order topological semi-metals~\cite{C1}. The quantization of fractional corner number of photons can be potentially used to achieve accurate topological light splitting~\cite{HOTIC}. Moreover, the coexistence of different dimensional topological boundary states can serve as a basis for designing topological switch circuits between crystalline insulators and higher-order topological insulators~\cite{C2}. In addition, our experimental realization which only relies on dielectric materials supports potential applications in developing lower-loss topological dielectric resonator antennas~\cite{C3}. Due to the scaling invariance in PC, the implementation can be minimized and the working frequency can be improved up to optical frequency by using micromachining technology~\cite{C4}. Finally, our findings may support explorations in designing topological point source lasers by introducing gain and loss in materials~\cite{C5,C6}, topological light-trappings~\cite{TRAP} and photonic chips with multi-photon quantum states~\cite{C7,C8,C9}.

{\sl Acknowledgments}.
This work was financially supported by National Key R\&D Program of China (Grant Nos. 2018YFA0306202 and 2017YFA0303702), and National Natural Science Foundation of China (Grant Nos. 11625418, 51732006, 11674166 and 11774162). B.X thanks Xueyi Zhu for the useful discussions on the simulations.

\end{document}